\documentclass[final,5p,times,twocolumn,authoryear]{elsarticle}

\makeatletter
\def\ps@pprintTitle{%
 \let\@oddhead\@empty
 \let\@evenhead\@empty
 \let\@oddfoot\@empty
 \let\@evenfoot\@empty
}
\makeatother
\usepackage{algorithmic}
\usepackage{graphicx}
\usepackage{algorithm,algorithmic}
\usepackage{hyperref}
\usepackage{textcomp}
\usepackage{graphicx}
\usepackage{subcaption}
\usepackage[utf8]{inputenc}
\usepackage[T1]{fontenc}
\usepackage{mathtools}
\usepackage{xcolor}
\usepackage{colortbl}
\usepackage{color}
\usepackage{url} 

\usepackage{amsmath,amssymb,amsfonts}
\usepackage{algorithmic}
\usepackage{graphicx}
\usepackage{algorithm,algorithmic}
\usepackage{hyperref}
\hypersetup{hidelinks=true}
\usepackage{textcomp}
\DeclareUnicodeCharacter{03B2}{\ensuremath{\beta}}

\renewcommand{\cite}[1]{\citep{#1}}

\begin{document}

%\preprint{APS/123-QED}

\begin{frontmatter}



\title{Alz-QNet: A Quantum Regression Network for Studying Alzheimer's Gene Interactions}

\if 0
\author[first]{Author name}
\affiliation[first]{organization={University of the Moon},%Department and Organization
            addressline={}, 
            city={Earth},
            postcode={}, 
            state={},
            country={}}
\fi 
\author[first]{Debanjan Konar}
\author[second]{Neerav Sreekumar}
\author[third]{Richard Jiang}
\author[first]{Vaneet Aggarwal}
\affiliation[first]{Purdue University, USA}\affiliation[second]{Indian Institute of Technology Tirupati, India}
\affiliation[third]{Lancaster University, UK}
\tnotetext[pubnote]{This is the author version of the work published in Computers in Biology and Medicine
Volume 196, Part C, September 2025, 110837.  DOI: \href{https://doi.org/10.1016/j.compbiomed.2025.110837}{10.1016/j.compbiomed.2025.110837}}

%\begin{indented}
%	\item[]August 2017 (minor update March 2024)
%\end{indented}
% \author{\IEEEauthorblockN{1\textsuperscript{st} Neerav Sreekumar}\\%and 6\textsuperscript{th} Vaneet Aggarwal}
% \IEEEauthorblockA{\textit{Department of Chemical Engineering,\\ Indian Institute of Technology, Tirupati}} \\
% \and\IEEEauthorblockN{2\textsuperscript{nd} Debanjan Konar and 3\textsuperscript{rd} Vaneet Aggarwal}\\
% \IEEEauthorblockA{\textit{Purdue Quantum Science and Engineering Institute,} \\
% \textit{School of Industrial Engineering, Purdue University, West Lafayette, USA}\\
% % %\textit{Laboratoire I3S, University Cote d'Azur, France} \\
% \textit{dkonar@purdue.edu}
% }}
% \author{\IEEEauthorblockN{1\textsuperscript{st} Neerav Sreekumar}
% \IEEEauthorblockA{\textit{Department of Chemical Engineering,} \\
% \textit{Indian Institute of Technology, Tirupati, India}}
% %\textit{Purdue University, West Lafayette, USA}\\
% \and
% \IEEEauthorblockN{2\textsuperscript{nd} Debanjan Konar and 3\textsuperscript{rd} Vaneet Aggarwal}
% \IEEEauthorblockA{\textit{Purdue Quantum Science and Engineering Institute (PQSEI), and } \\
% \textit{and School of Industrial Eng., Purdue University, West Lafayette, USA} \\
% }
% }
%\textit{Görlitz, Germany}, a.cangi@hzdr.de}}

%\maketitle

\begin{abstract}
Understanding the molecular-level mechanisms underpinning Alzheimer's disease (AD) by studying crucial genes associated with the disease remains a challenge. Alzheimer's, being a multifactorial disease, requires understanding the gene-gene interactions underlying it for theranostics and progress. In this article, a novel attempt has been made using a quantum regression to decode how some crucial genes in the AD Amyloid Beta Precursor Protein ($APP$), Sterol regulatory element binding transcription factor 14 ($FGF14$), Yin Yang 1 ($YY1$), and Phospholipase D Family Member 3 ($PLD3$) etc. become influenced by other prominent switching genes during disease progression, which may help in gene expression-based therapy for AD. Our proposed Quantum Regression Network (Alz-QNet) introduces a pioneering approach with insights from the state-of-the-art Quantum Gene Regulatory Networks (QGRN) to unravel the gene interactions involved in AD pathology, particularly within the Entorhinal Cortex (EC), where early pathological changes occur. Using the proposed Alz-QNet framework, we explore the interactions between key genes ($APP$, $FGF14$, $YY1$, $EGR1$, $GAS7$, $AKT3$, $SREBF2$, and $PLD3$) within the CE microenvironment of AD patients, studying genetic samples from the database $GSE138852$, all of which are believed to play a crucial role in the progression of AD. Our investigation uncovers intricate gene-gene interactions, shedding light on the potential regulatory mechanisms that underlie the pathogenesis of AD, which help us to find potential gene inhibitors or regulators for theranostics.
\end{abstract}


\end{frontmatter}

\if 0
\begin{IEEEImpStatement}
The Alz-QNet paper introduces a novel quantum regression network for analyzing Alzheimer's disease (AD) gene interactions, offering insights into the molecular mechanisms of AD. The model reduces computational complexity by 50\% compared to traditional quantum gene regulatory networks, enabling efficient exploration of high-dimensional genomic data. Alz-QNet highlights regulatory relationships among crucial AD-related genes such as APP, SREBF2, and PLD3, and uncovers key interactions within the entorhinal cortex, a primary site for early AD pathology. This innovative approach not only advances understanding of AD’s multifactorial nature but also provides a computational framework to identify potential therapeutic targets. The interdisciplinary integration of quantum computing and genomic analysis sets a new benchmark for precision in studying neurodegenerative diseases, promising significant strides in personalized medicine and computational biology.
\end{IEEEImpStatement}
%\footnote{\scriptsize{The Pytorch code for our Alz-QNet implementation is available at \url{https://anonymous.4open.science/r/QuantumGRN-E255}.}}
\fi

%\maketitle

\section{Introduction}
\label{chapter:intro}

Alzheimer's disease (AD) presents a formidable challenge in healthcare, characterized by progressive cognitive decline and neurodegeneration~\cite{grubman2019}. The accumulation of peptides of Amyloid Beta ($A\beta$) from $APP$ is a crucial factor in the pathology of AD~\cite{wang2018}. Despite extensive research efforts over decades, the intricate mechanisms underlying AD remain elusive, impeding the development of effective therapeutic interventions~\cite{palop2010}. The hypothesis of the amyloid cascade, which attributes neurotoxicity and neuronal loss to the accumulation of $A\beta$ peptides, has been a prominent theory but has faced limitations in translating it into successful treatments~\cite{nowak2006}. Clinical trials targeting $A\beta$ accumulation have produced disappointing results, underscoring the multi-factorial nature of AD pathogenesis influenced by factors such as Reactive Oxygen Species (ROS) and ferroptosis~\cite{vossel2017}. Mounting evidence suggests that $A\beta$ deposition may be a downstream consequence rather than the primary driver of neurodegeneration, necessitating a reevaluation of therapeutic strategies and a deeper understanding of the molecular underpinnings of AD. {The failure of the Amyloid Cascade hypothesis proved why AD is a multifactorial disease by debunking the myth that the gene that causes upregulation of amyloid beta, APP, is the primary driver of AD progression. Thus, the requirement to study a multigene model in the EC environment for potential gene control-based therapy has great potential.}\\
Gene Regulatory Networks (GRN)~\cite{grn2008, david2005} are complex systems that govern gene expression and cellular processes. Understanding the dynamics of GRN is essential to elucidate the molecular mechanisms underlying physiological functions and disease states. In AD, the entorhinal cortex is particularly significant due to its early involvement in disease progression~\cite{selko2016}. However, deciphering AD-related GRN presents challenges, which require innovative computational approaches to integrate multi-omics data and reveal the regulatory landscape of AD-associated genes~\cite{hoesen1991}. Correlation- and regression-based methods are commonly employed techniques for GRN inference because of their computational efficiency. These methods typically compute correlation or regression coefficients for gene pairs based on the total number of cells in the dataset. However, they have limitations, as they treat gene pairs independently, failing to fully capture complex expression patterns by incorporating additional layers of information.\\
Quantum Computing (QC) harnesses the principles of quantum mechanics to perform computations beyond the capabilities of classical computers~\cite{laura2022}. Unlike classical bits, which can only exist in a state of either $0$ or $1$, quantum bits or qubits can exist in superpositions of these states, enabling exponentially greater computational capacity and effective computations~\cite{konar2023}. This exponential scaling opens avenues for solving computationally intractable problems in genomics~\cite{pal2023}. For example, the human genome is given by $3$ billion base pairs, which can be represented by \(10^{10}\) classical bits, which are equivalent to $34$ qubits (\(2^{n}\) possible states for each). Building upon the foundations of Quantum Machine Learning (QML) seeks to leverage quantum algorithms and hardware to enhance traditional machine learning techniques~\cite{jacob2018}. QML offers the promise of accelerated learning and improved performance in large datasets by exploiting quantum parallelism and entanglement~\cite{konar2022}. Moreover, QML can potentially address challenges such as feature selection, dimensionality reduction, and pattern recognition, extending the applicability of machine learning to complex scientific domains~\cite{jacob2018, pal2023, konar2023_rqnn, konar2023_dsqnet}. \\
However, the main problem lies in the computational expense and complexity of quantum circuits, considering the costs of gates like controlled rotation, which are extensively used in QML (regression) circuits, which do not entirely make quantum computing an economically feasible option for data science in the short term~\cite{konar2023_trqnet}. The traditional method of Quantum Gene Regulatory Networks (QGRN)~\cite{roman2022} suffers from being computationally very expensive to implement, as the quantum circuit with $n$ qubits requires $n(n-1)$ controlled rotation gates and $n$ rotation gates. \\
Recognizing these limitations, our Alz-QNet model harnesses the computational power of QML to explore high-dimensional genomic data to reveal intricate gene-gene interactions that contribute to the pathogenesis of AD~\cite{sarn2018}. In the midst of the intricate molecular mechanisms and unresolved queries in AD research, the integration of QML and GRN analysis holds promise to unravel the complexities of AD pathogenesis. Using quantum regression networks (Alz-QNet) inspired by GRN, the objective is to uncover the regulatory dynamics controlling key genes associated with AD, such as $APP$, $SREBF2$, and $EGR1$. Through this interdisciplinary effort, the aim is to advance the understanding of AD pathogenesis and lay the foundation for innovative therapeutic interventions targeting dysregulated gene networks~\cite{lamb1998}.\\The primary contributions of our work are summarized in the following. 
\begin{enumerate}
    \item In this study, {we optimized Quantum Gene Regulatory Networks (QGRN)~\cite{roman2022}} to address the computational complexity and expense associated with the construction of variational quantum circuits for larger datasets. Our proposed Alz-QNet requires $\frac{n (n-1)}{2}$ controlled rotation gates over $n(n-1)$ for the QGRN model, demonstrating that a single controlled rotation gate between 2 qubits is sufficient to measure the interaction. 
    \item {In our Alz-QNet circuit, the reduction of $C_{R_Y}$ gates by half compared to the traditional QGRN~\cite{roman2022}, optimizes the number of variational parameters, $\theta_{x,y}$ between two qubits $x$ and $y$. We observed that the value of $\theta_{x,y}$ = $\theta_{y,x}$. It enhances the scalability potential of our proposed Alz-QNet.}  
    \item The study also aimed to uncover how $APP$ interacts with other key genes in AD, potentially offering information on control-based therapies for the gene expression of the disease.
    \item Furthermore, beyond $APP$, the research used single nucleus RNA sequencing of the entorhinal cortex to explore less explored genes such as $YY1$, $SREBF2$, $PLD3$, $GAS7$, and $EGR1$.
\end{enumerate}
{\bf Motivation: }
\label{motiv}
% Traditional approaches to QGRN~\cite{roman2022} are computationally demanding, necessitating a substantial number of controlled rotation gates and rotation gates for each qubit in the quantum circuit. Our model specifically focuses on investigating eight genes linked to AD and seeks to decrease computational complexity by reducing the number of gates $C_{R_Y}$ by half through a bypass mechanism. This novel strategy enables the exploration of classical GRN with a reduced computational load, particularly within the realm of AD-related research. The forthcoming Section~\ref{Architecture:RQNN} elaborates on this methodology, encompassing quantum circuit design and the innovative method to examine gene interactions within AD pathology.
{QML research in general has demonstrated better performance in term of generalization, expressivity, privacy and robustness~\cite{konar2023_rqnn, konar2023_dsqnet, caro2022}. It presents several notable advantages over the classical GRN reconstruction methods~\cite{roman2022}. The QML algorithms utilized in GRN research leverage quantum superposition and entanglement, which allows them to capture complex, non-linear relationships between genes that classical algorithms may overlook. Furthermore, QML is better equipped to address the curse of dimensionality, which is especially pertinent given the high-dimensional nature of genomic data.}

\section{Various Gene Types with Alzheimer's Diseases}
\label{sec:gene_types}

Comprehending the regulatory connections among pivotal genes associated with AD, such as $APP$, $SREBF2$, and other relevant genes, is essential for deciphering the molecular mechanisms driving AD pathogenesis. These genes are fundamental in synaptic function, lipid metabolism, and neuronal viability, and their dysregulation is associated with AD-related neuropathological processes, including $A\beta$ accumulation, tau hyperphosphorylation, and synaptic dysfunction~\cite{bottero2021}.\\
% Alzheimer's disease (AD) presents one of the most pressing challenges in modern healthcare, with its devastating impact on individuals and societies worldwide. Central to its pathology is the dysregulation of multiple genes, each playing crucial roles in various aspects of neuronal function, metabolism, and survival. Understanding how these genes interact and influence AD progression is paramount for developing effective therapeutic strategies.\\
At the forefront of AD research lies the $APP$, a gene pivotal to the disease's pathogenesis. Anomalies in $APP$ processing lead to the generation of amyloid beta peptides, whose accumulation forms insoluble plaques - a hallmark of AD pathology~\cite{selko2016, zhang2011}. This neurotoxic cascade disrupts synaptic function, induces oxidative stress, and triggers inflammatory responses, ultimately resulting in synaptic dysfunction and neuronal loss. Continuing investigations actively explore strategies to modulate $APP$ metabolism, prevent $A\beta42$ aggregation, or improve $A\beta42$ clearance as potential therapeutic interventions for AD~\cite{julia2017}. \\
However, $APP$ is one component of the intricate puzzle of AD. Other genes, such as $AKT3$, $SREBF2$, $YY1$, $GAS7$, $EGR1$, $FGF14$, and $PLD3$, also play crucial roles in the pathogenesis of AD. For example, dysregulation of $AKT3$ signaling has been associated with insulin resistance, a common feature of AD pathology that exacerbates neurodegeneration. Studies suggest that activated $AKT3$ promotes neuronal survival by inhibiting Glycogen Synthase Kinase-3 $\beta$ ($GSK-3 \beta$), a pro-apoptotic protein, and enhances adult neurogenesis in neural stem cells (NSC). Conversely, dysregulated $AKT3$ signaling may worsen neuronal vulnerability and contribute to AD pathogenesis, particularly in conditions like insulin resistance~\cite{huang2012}. Therefore, exploring the molecular mechanisms underlying $AKT3$ dysregulation in AD provides valuable information on potential therapeutic targets to mitigate neurodegeneration and cognitive decline in individuals with AD~\cite{pluta2020,razani2021}. \\
$SREBF2$, a key regulator of lipid metabolism, interacts with pathways involving crucial genes in AD, impacting processes such as amyloid-beta production, transcriptional regulation, and insulin signaling. Dysregulation of $SREBF2$ signaling is associated with neurodegeneration, exacerbating amyloid beta accumulation, synaptotoxicity, and memory deficits in neuronal cells~\cite{barb2018}. Furthermore, disrupted $SREBF2$ signaling affects cholesterol homeostasis in brains with AD, preventing the clearance of plaques $A\beta$ and increasing oxidative stress in neuronal cells. Recent research has highlighted the role of $SREBF2$ in AD, demonstrating significant changes in its nuclear translocation and activation patterns in AD brains and relevant animal models. Specifically, reduced nuclear translocation of mature $SREBF2$ ($mSREBP2$) is observed in AD brains, with tau modifications associated with these alterations~\cite{yue2022,wang2018}. \\
% $YY1$, a transcription factor, plays a crucial role in regulating genes essential for neuronal survival and synaptic function in AD. Dysregulated activity $YY1$ has been associated with increased $APP$, contributing to plaque formation in the brain of AD. Furthermore, dysregulation of $YY1$ activity has been associated with elevated expression of $BACE1$, leading to enhanced amyloid beta production and plaque formation in AD. Furthermore, $YY1$ influences genes such as $Fuz$ and $APH1A$, which impact neurodegenerative processes in AD.\\
The involvement of $YY1$ in regulating the $Fuz$ gene adds another layer of significance to its role in AD research. Aberrant $YY1$ activity can result in excessive methylation of the $Fuz$ gene promoter, leading to reduced transcription. This alteration affects the polarity of the planar cells and subsequent cell stability, which are crucial for neuronal health. The elevated $Fuz$ transcript levels observed in individuals with AD pathology suggest that $YY1$-mediated modifications of the $Fuz$ gene may contribute to neuronal apoptosis and neurodegeneration in AD. Thus, unraveling the interplay between $YY1$ and the $Fuz$ gene provides information on additional mechanisms underlying AD pathogenesis, offering potential avenues for therapeutic intervention targeting $YY1-Fuz$ interactions to mitigate neuronal loss and cognitive decline in AD. Targeting $YY1$ expression or activity could hold therapeutic potential for alleviating AD pathology, making it a significant focus in the pursuit of effective AD treatments~\cite{wein2017, nowak2006}.\\
$GAS7$, $EGR1$, $FGF14$, and $PLD3$ each play distinct roles in AD pathogenesis, influencing processes such as tau phosphorylation, synaptic plasticity, and lipid metabolism. Dysregulated expression of $GAS7$ in neurons has been associated with altered microtubule transport proteins, potentially leading to tau dysregulation and increased susceptibility to AD development. Recent research has shed light on the involvement of $GAS7$ in neuronal maturation and morphogenesis, further implicating its role in the progression of AD~\cite{cunn2020}. $GAS7$ expression promotes the formation of dendrite-like processes and filopodia projections in neuronal cells, enhancing neurite outgrowth and microtubule bundling. These findings suggest that $GAS7$ governs neural cell morphogenesis by coordinating actin filaments and microtubules, thereby influencing neuronal maturation and potentially impacting AD progression. The intricate interplay among these genes unveils a complex network of molecular events driving AD progression~\cite{gotoh2013}. $EGR1$'s modulation of acetylcholinesterase mRNA and protein levels indicates its significant contribution to alterations in acetylcholine signaling observed in AD, where acetylcholine depletion is prominent. Furthermore, $EGR1$'s regulation of $miRNA-132$, impacting the nucleus basalis of Meynert rich in acetylcholine, underscores its role in AD-related neurotransmitter dysregulation~\cite{hu2019}. Recent studies have provided promising information on the therapeutic potential of $EGR1$ in AD. Silencing $EGR1$ in AD mouse models reduces tau phosphorylation, decreases amyloid-beta pathology, and improves cognition~\cite{hutt1998}. \\
$EGR1$ regulates tau phosphorylation and amyloid synthesis by influencing the activities of $Cdk5$ and $BACE-1$, respectively, suggesting its potential as a therapeutic candidate for the treatment of AD~\cite{qin2017}. Conversely, dysfunction in sodium channel signaling due to $FGF14$ deficiency has been linked to neurological disorders, such as schizophrenia~\cite{hsu2017}. In particular, modulation of sodium channel signaling $FGF14$' can affect amyloid beta pathology, with $PPAR-\gamma$ agonists showing promise by phosphorylating $FGF14$ and modulating sodium channel signaling. This suggests a potential role for $FGF14$ as a therapeutic target in the management of neuronal dysfunction and memory loss observed in early AD. Additionally, $FGF14$ demonstrates neuroprotective effects by inhibiting $MAPK$ signaling, highlighting its potential as a therapeutic agent for neurodegenerative conditions. Its complex interactions with voltage-gated sodium channels at the axonal initial segment influence neuronal excitability, synaptic transmission, and neurogenesis, affecting cognitive and affective behavioral outcomes~\cite{jessica2017}. In translational studies, $FGF14$ has been increasingly associated with diseases related to cognitive and affective domains, including neurodegeneration, indicating its involvement as a converging node in the etiology of complex brain disorders, further emphasizing its potential significance in AD pathogenesis~\cite{wang2021}.\\
$PLD3$'s involvement in the formation of amyloid plaque-associated axonal spheroids highlights its role in the dysfunction of the neural network in AD~\cite{wang2015}. Mechanically, $PLD3$ encodes a highly concentrated lysosomal protein in axonal spheroids, with its overexpression leading to spheroid enlargement and exacerbated axonal conduction blockades. In contrast, deletion of $PLD3$ reduces the size of the spheroid and improves the function of the neural network. Targeted modulation of endolysosomal biogenesis mediated by $PLD3$ in neurons presents a promising avenue to reverse axonal spheroid-induced neural circuit abnormalities in AD, independent of amyloid removal. In AD brains, suppressing inappropriate PLD signaling has shown the potential to enhance synaptic resilience and decelerate cognitive decline, offering therapeutic advantages in AD management~\cite{yuan2022}.\\
Studying the gene regulation of all eight genes collectively provides a comprehensive understanding of AD pathophysiology. By elucidating the dynamic interactions among these genes, researchers can uncover common underlying mechanisms and novel therapeutic targets for AD~\cite{yuan2022}.

%{\bf \color{red} Need formatting fixes in this section. There seems only 1 subsection and then 1. ... are unclear}
\subsection{Single-Nucleus RNA Sequencing}
\label{snRNA}
Single-nucleus RNA sequencing ($snRNA-seq$) is a powerful technique to analyze gene expression patterns at the single-cell level. Unlike traditional RNA sequencing methods, which require intact cells, $snRNA-seq$ enables analysis of gene expression of individual nuclei extracted from tissues, including complex tissues such as the brain~\cite{wu2018}. The snRNA-seq process involves several steps:
\begin{enumerate}
    \item Nuclei Isolation: Tissue samples are dissociated to release individual nuclei while preserving RNA integrity.
\item Library Preparation: RNA is extracted from isolated nuclei, and cDNA libraries are generated by reverse transcription.
\item Sequencing: cDNA libraries are sequenced using high-throughput sequencing platforms, generating millions of short reads corresponding to RNA transcripts.
\item Data Analysis: Bioinformatics tools are used to align sequencing reads to a reference genome, quantify gene expression levels, and perform downstream analyzes, such as identifying cell types and characterizing gene regulatory networks.
\item Studying Gene Expression: $snRNA-seq$ provides information on gene expression heterogeneity within cell populations and allows researchers to identify rare cell types and transcriptomic changes associated with various biological processes and disease states. By profiling gene expression at the single-cell level, $snRNA-seq$ enables the discovery of novel cell types, regulatory pathways, and biomarkers with high resolution and sensitivity~\cite{jovic2022}. 
\end{enumerate}

\section{Quantum Computing Theory}
\label{sec:qc}

\subsection{Qubits and Basis States}
In quantum computing, the basic unit is \emph{qubit}. A qubit can exist in a superposition of the states $|0\rangle$ and $|1\rangle$, represented as~\cite{rieffel2000}:
\begin{equation}
|\psi\rangle = \alpha |0\rangle + \beta |1\rangle\,,
\end{equation}
where $\alpha$ and $\beta$ are complex numbers such that $|\alpha|^2$ + $|\beta|^2 = 1$.
The states $|0\rangle$ and $|1\rangle$ are the computational basis states.

\subsection{Quantum Gates}
Quantum gates manipulate qubits through unitary operations. Here are some basic quantum gates and their operations~\cite{rieffel2000}:
\begin{itemize}
\item NOT Gate: The NOT ($X$) gate flips the state of a qubit as follows~\cite{konar2023_sqnn}:
\begin{equation}
X = \begin{pmatrix}
0 & 1 \\
1 & 0
\end{pmatrix}\,,
\end{equation}
\begin{equation}
X |0\rangle = |1\rangle, \quad X |1\rangle = |0\rangle\ .
\end{equation}
\item Hadamard Gate: The Hadamard ($H$) gate  creates a superposition of the states $|0\rangle$ and $|1\rangle$~\cite{konar2023_sqnn}:
\begin{equation}
H = \frac{1}{\sqrt{2}} \begin{pmatrix}
1 & 1 \\
1 & -1
\end{pmatrix}\,,
\end{equation}
\begin{equation}
H |0\rangle = \frac{1}{\sqrt{2}} (|0\rangle + |1\rangle), \quad H |1\rangle = \frac{1}{\sqrt{2}} (|0\rangle - |1\rangle)\ .
\end{equation}
\item Controlled-NOT Gate:
The Controlled-NOT ($C_X$) gate flips the target qubit if the control qubit is in the state $|1\rangle$~\cite{konar2023_sqnn}:
\begin{equation}
C_X = \begin{pmatrix}
1 & 0 & 0 & 0 \\
0 & 1 & 0 & 0 \\
0 & 0 & 0 & 1 \\
0 & 0 & 1 & 0
\end{pmatrix}\ .
\end{equation}
The action of a $C_X$ gate on the basis states:
\begin{equation}
C_X |00\rangle = |00\rangle, \quad C_X |01\rangle = |01\rangle\ .
\end{equation}
\begin{equation}
C_X |10\rangle = |11\rangle, \quad C_X|11\rangle = |10\rangle\ .
\end{equation}
\item Rotation Gate:
The rotation gate around the Y-axis ($R_Y$) rotates the qubit state by an angle $\theta$~\cite{konar2023_sqnn}:
\begin{equation}
R_Y(\theta) = \begin{pmatrix}
\cos(\frac{\theta}{2}) & -\sin(\frac{\theta}{2}) \\
\sin(\frac{\theta}{2}) & \cos(\frac{\theta}{2})
\end{pmatrix}\ .
\end{equation}
\item Controlled-$R_Y$ Gate:
The controlled-$R_Y$ ($C_{R_Y}$) gate rotates around the $Y$-axis on the target qubit if the control qubit is in the state $|1\rangle$. This can be constructed using a $C_X$ gate and $R_Y$ gates as follows~\cite{konar2023_sqnn}:
%The matrix representation of the controlled $R_Y$ gate is:
\begin{equation}
C_{R_Y}(\theta) = \begin{pmatrix}
1 & 0 & 0 & 0 \\
0 & 1 & 0 & 0 \\
0 & 0 & \cos(\frac{\theta}{2}) & -\sin(\frac{\theta}{2}) \\
0 & 0 & \sin(\frac{\theta}{2}) & \cos(\frac{\theta}{2})
\end{pmatrix}\ .
\end{equation}
This operation can be decomposed into a sequence involving a $C_X$ gate and $R_Y$gates:
\begin{equation}
C_{R_Y}(\theta) = (I \otimes R_Y(\frac{\theta}{2})) \cdot C_X \cdot (I \otimes R_Y(-\frac{\theta}{2})) \cdot C_X \ .
\end{equation}
Here, $I$ is the identity matrix. The standard basis state table for controlled rotation gates is provided in Table~\ref{table:cr_y_mapping}
\end{itemize}
%In a quantum circuit, the implementation looks like this:
% \begin{equation}
% \Qcircuit @C=1em @R=.7em {
% & & & \ctrl{1} & \gate{R_Y(\frac{\theta}{2})} & \ctrl{1} & \gate{R_Y(-\frac{\theta}{2})} & \qw \\
% & & & \targ & \qw & \targ & \qw & \qw 
% }
% \end{equation}
%\subsection{The standard basis state table for controlled rotation gates}
\begin{table}[h]
\caption{Mapping basis states using a $C_{R_Y}(\theta)$ gate}
\centering
\begin{tabular}{|c|c|}
\hline
\textbf{Basis state $|x\rangle$} & \textbf{$C_{R_Y}(\theta) |x\rangle$} \\ \hline
$|00\rangle$ & $|00\rangle$ \\ \hline
$|01\rangle$ & $|01\rangle$ \\ \hline
$|10\rangle$ & $\cos\left(\frac{\theta}{2}\right) |10\rangle + \sin\left(\frac{\theta}{2}\right) |11\rangle$ \\ \hline
$|11\rangle$ & $-\sin\left(\frac{\theta}{2}\right) |10\rangle + \cos\left(\frac{\theta}{2}\right) |11\rangle$ \\ \hline
\end{tabular}
\label{table:cr_y_mapping}
\end{table}

\section{Quantum Regression Network Architecture}
\label{Architecture:RQNN}

\begin{figure*}[htbp]
	\centering
	\includegraphics[scale=0.5]{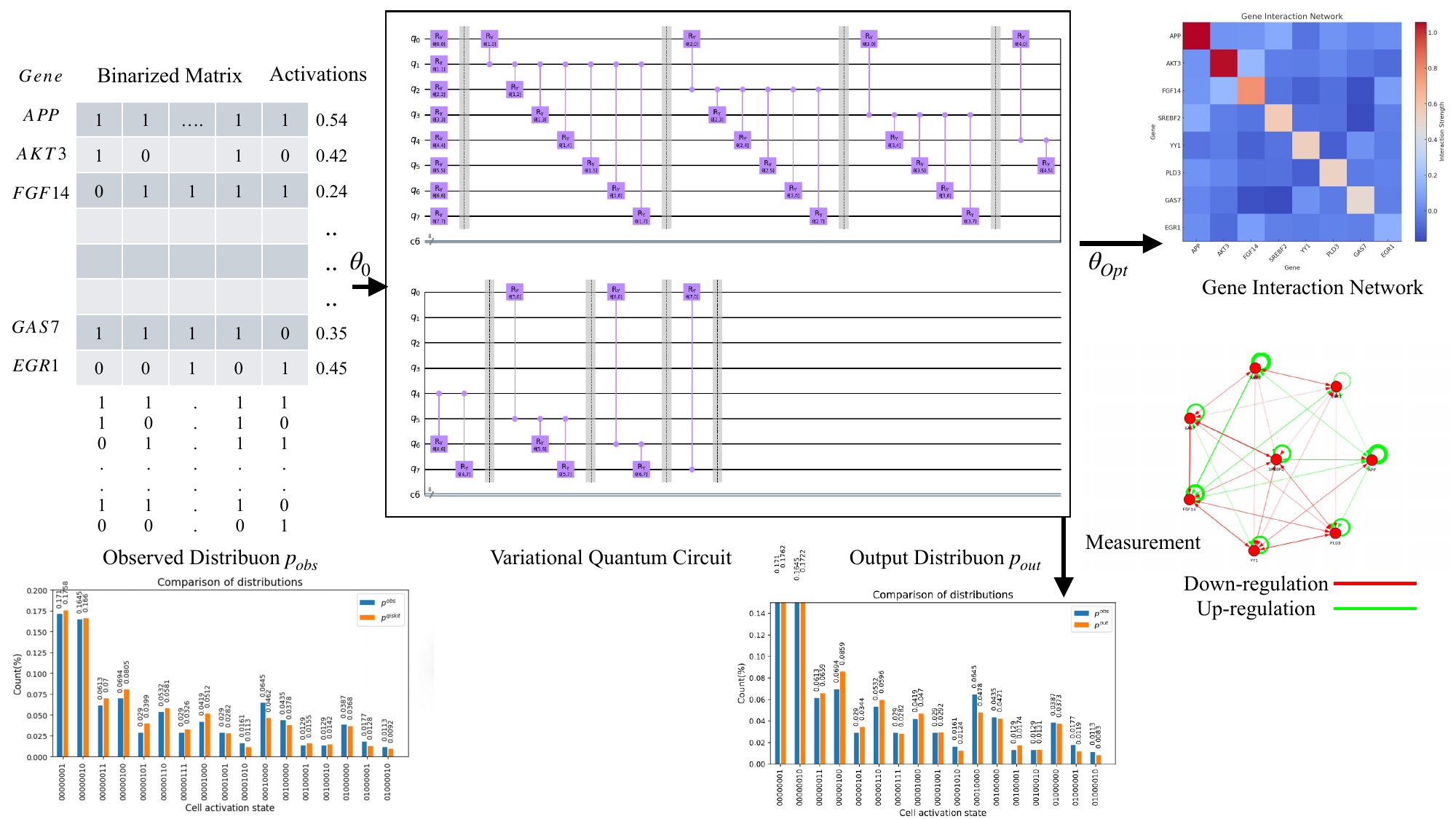} 
    \caption{{A Quantum Regression Network (Alz-QNet) relying on Variational Quantum Circuit (VQC) to study Alzheimer's gene interactions (The circuit is continued from top to bottom).}}
\label{fig:QGRN}
\end{figure*}
\begin{figure*}[htbp]
	\centering
	\includegraphics[scale=0.4]{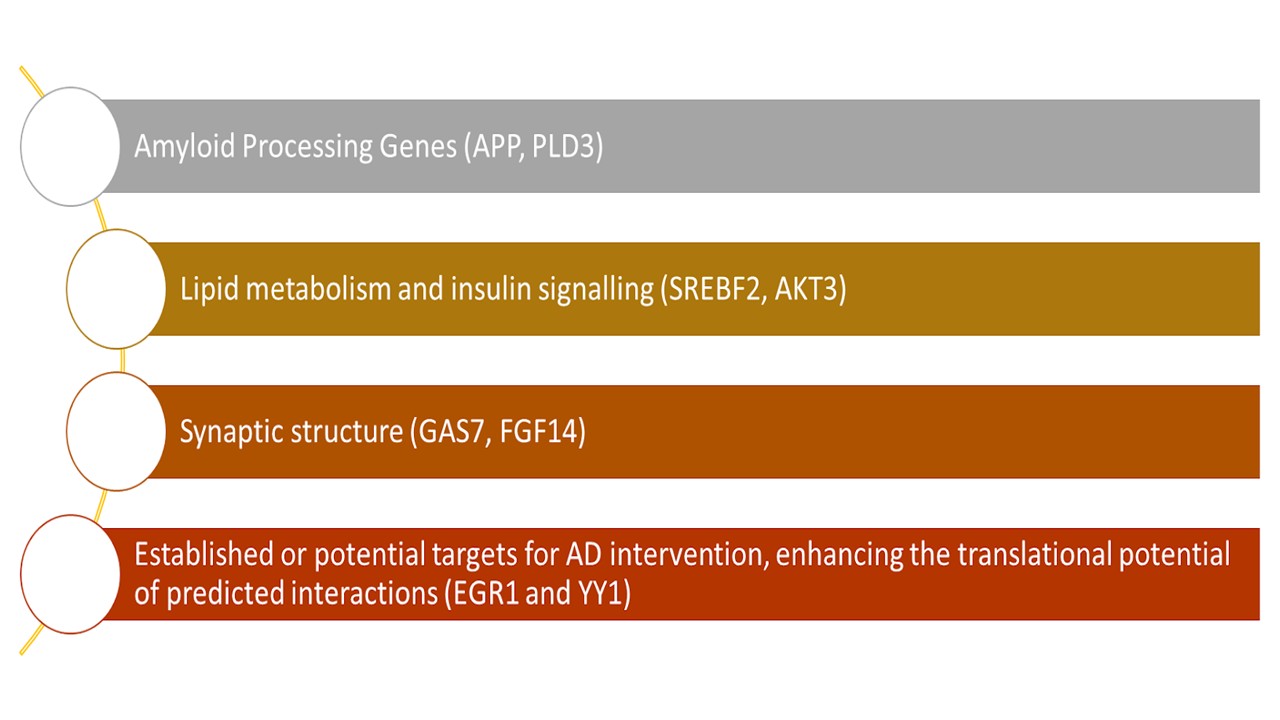} 
    \caption{\textcolor{black}{Pathway Map for the criteria of gene selection in Alz-QNet}.}
\label{fig:QGRN}
\end{figure*}

The proposed Alz-QNet model centered on studying the 8 genes specific to Alzheimer's disease by significantly reducing the cost of the $C_{R_Y}$ gates through a bypass mechanism, thereby decreasing their number by half to analyze Gene Regulatory Networks (GRN)~\cite{grn2008} with reduced computational complexity. Inspired by the Quantum Gene Regulatory Networks (QGRN)~\cite{roman2022}, our proposed Alz-QNet, as shown in Fig.~\ref{fig:QGRN}, leverages the quantum entanglement quantum computing offers to explore gene regulatory relationships. \\ In the proposed Alz-QNet model, each qubit represents a gene and initializes to phase $0$. The Alz-QNet is structured into two sections: the encoder and regulation layers. The encoder layer translates $snRNA-seq$ data into a superposition state, while the regulation layers entangle qubits to model gene-gene interactions within the quantum framework. Through these layers, we construct an $8\times 8$ matrix, with an initial unknown value of $\theta_{x,y}$ in the $C_{R_Y}$ gates entangling $2$ qubits, where $x$ represents the control qubit (gene) and $y$ represents the target qubit (gene). The optimized values of each $\theta$ in the matrix correspond to the strength of gene interaction. Traditional Laplace smoothing and the gradient descent algorithm are used for the optimization process to minimize a loss function based on Kullback-Leibler (KL) divergence~\cite{naga2009}.\\
Our Alz-QNet model utilizes a $C_{R_Y}$ gate to establish connections between each pair of qubits in the regulation layers, simulating the regulatory relationships between two genes. The rotation angle of the $C_{R_Y}$ gate signifies the strength of the interaction between the control gene and the target gene. Following optimization, these rotation angles are parameterized and translated into the adjacency matrix to construct a GRN. %Throughout our investigation, we initially assumed that a larger rotation angle indicated a stronger interaction. However, we discovered that this assumption is not always accurate. To exemplify this concept, we present an illustration in Fig.~\ref{fig:graph1}. 
A primary unit circuit is depicted in Fig.~\ref{fig:graph1}, initialized in the $|00\rangle$ state. This circuit comprises a control qubit ($1^{st}$ qubit, rotated with a $R_Y$ gate at an angle $\phi_1$), a target qubit ($2^{nd}$ qubit, rotated with a $R_Y$ gate at an angle $\phi_2$) and a $C_{R_Y}$ gate with a rotation angle $\theta$.\\
% Correlation-based and regression-based methods are the most widely used techniques for GRN inference due to their computational efficiency. These methods typically calculate correlation or regression coefficients for gene pairs using the total number of cells in the data. However, they have limitations as they handle gene pairs independently, without fully capturing complex expression patterns by incorporating additional levels of information. The relationship between any two genes is measured using a single summary statistic, such as a correlation or regression coefficient, which remains unaffected by the total number of cells. Therefore, increasing the number of cells has little impact on these coefficients.\\
% Another limitation is that these coefficients are calculated for gene pairs in isolation, ignoring the expression values of other genes in the same biological system. This can result in biased coefficients that do not accurately reflect true interactions. Although adjustments can be made, they are limited since modelling all-to-all interactions is challenging.\\
% In our Alzheimer's disease model (GSE 138852), we are studying gene regulatory relationships among 8 specific genes. Our approach aims to provide a more comprehensive understanding of these interactions by considering the complexities and nuances of gene expression patterns specific to this condition.
%\subsection{Measuring the Output Register of the qscGRN Model}
We measured the output register to obtain the output distribution $p_{out}$ of the basis states. The probability of a particular state in $p_{out}$ was set to $0$, and the remaining distribution was rescaled to sum to $1$.
%\subsection{Smoothing $p_{\text{obs}}$ and $p_{\text{out}}$}
Laplace smoothing was applied to reshape $p_{\text{obs}}$ and $p_{\text{out}}$ into smoothed distributions $\tilde{p}_{\text{obs}}$ and $\tilde{p}_{out}$, respectively. These smoothed distributions are computed as follows:
\begin{equation}
\tilde{p}(x) = \frac{p(x) + \alpha}{N + \alpha \cdot |X|}\,,
\end{equation}
where $\alpha$ is the smoothing parameter (typically set to $1$), $N$ is the total number of occurrences in the distribution, and $|X\rangle$ is the size of the distribution's support. In other words, $\tilde{p}$ represents the original distribution after smoothing.

\subsection{Loss Function}

The loss function, comprising the KL divergence ($\mathcal{L}_{KL}$)~\cite{naga2009} and a constraint term ($\mathcal{L}_c$), is defined as follows:
\begin{equation}
\mathcal{L}_{KL} = \sum p_{obs} \log \frac{p_{obs}}{p_{out}}\ .
\end{equation}
\begin{equation}
\mathcal{L}_c = \sum \left( \theta - \theta_0 \right)^2\,,
\end{equation}
where $\theta$ is the parameter in the proposed Alz-QNet model and $\theta_0$ is the initial parameter set. Thus, the total loss function was defined as:
\begin{equation}
\mathcal{L} = \mathcal{L}_{KL} + \lambda \mathcal{L}_c \,,
\end{equation}
where $\lambda$ is a dynamic coefficient that rescales $\mathcal{L}_c$ to the same order of magnitude as $\mathcal{L}_{KL}$. In summary, the term $\mathcal{L}_{KL}$ aligns the output distribution $p_{out}$ with the observed distribution $p_{obs}$, while the term $\mathcal{L}_c$ prevents any parameter $\theta$ from deviating significantly from $\theta_0$.\\
%\subsection{Optimization of the Parameter}
Optimization has been performed by iteratively minimizing the loss function until it has reached a threshold value of $2^n \times 10^{-4}$ using a modified gradient descent algorithm with a learning rate ($\eta$) of $0.05$. If this threshold is unmet, optimization has continued for a predefined number of iterations $\iota$. The parameter $\theta$ in iteration $\iota$ was updated as:
\begin{equation}
\theta^{(\iota+1)} = \theta^{(\iota)} - \eta \nabla \mathcal{L}^T\,,
\end{equation}
where $\nabla \mathcal{L}$ is the gradient of the loss function, ensuring $\theta$ remains a symmetric matrix. 
%as we noted that the value of $\theta_{a,b} = \theta_{b,a}$, hence proving the fact that only a single controlled rotation gate between 2 qubits is sufficient to measure the interaction, provided we manually equate $\theta_{a,b}$ = $\theta_{b,a}$ to construct the entire symmetric matrix. Our work is integrated with Qiskit—an open-source quantum computing library that simulates a noisy quantum circuit using the Aer Simulator backend with default parameters.\\
In the proposed Alz-QNet, $\theta_{k,k}$ as the parameter for the $R_Y$ gate on the $k^{th}$ qubit in the $L_e$ layer, and $\theta_{k,p}$ for the $C_{R_Y},n$ gate with the $k^{th}$ qubit as control and the $p^{th}$ qubit as target in the $L_k$ layer of an $n$-qubit system. For our case where $n=8$, the layers are defined as follows:
\begin{equation}
L_e = R_Y(\theta_{7,7}) \otimes R_Y(\theta_{6,6}) \otimes \cdots \otimes R_Y(\theta_{1,1}) \otimes R_Y(\theta_{0,0})\,,
\end{equation}
and
\begin{equation}
\begin{split}
L_k = \prod_{i=0, i \ne k}^{7} C_{R_Y},n(\theta_{k,i})
= C_{R_Y},n(\theta_{k,7}) \\\otimes \cdots \otimes c- R_Y,n(\theta_{k,1}) \otimes C_{R_Y},n(\theta_{k,0})\ .
\end{split}
\end{equation}
In our Alz-QNet circuit, we have reduced the number of $C_{R_Y}$ gates by half compared to the traditional QGRN~\cite{roman2022}, optimizing the upper triangular matrix of $\theta_{x,y}$. We observed that the value of $\theta_{a,b}$ = $\theta_{b,a}$, demonstrating that a single controlled rotation gate between 2 qubits is sufficient to measure the interaction. By manually setting $\theta_{a,b}$ = $\theta_{b,a}$, we construct the entire matrix:
\begin{equation}
\theta =
\begin{pmatrix}
\theta_{0,0} & \theta_{0,1} & \cdots & \theta_{0,7} \\
\theta_{1,0} & \theta_{1,1} & \cdots & \theta_{1,7} \\
\vdots & \vdots & \ddots & \vdots \\
\theta_{7,0} & \theta_{7,1} & \cdots & \theta_{7,7}
\end{pmatrix}\ .
\end{equation}
In the modified constraint loss $L_c$, the parameter $\theta_0$ represents the initial rotation angles of the quantum gates, reflecting prior assumptions or biological priors. In our implementation, $\theta_0$ is initialized to $0$ for all non-diagonal elements in the matrix, thus having $0$ bias. All diagonal elements are having $\theta_0$ of $2 \arcsin(\sqrt{a_k})$, where $a_k$ is the activation ratio for the $k^{th}$ gene.\\
\textcolor{black}{Our Alz-QNet’s gate complexity scales as $O(\frac{n^2}{2})$ versus $O(n^2)$ in traditional QGRN—reducing gates from $9,900$ to $4,950$ for $100$ genes. Current hardware supports $50$-gene networks; $1,000$-gene networks become feasible with next-gen $1,000+$ qubit processors. The proposed Alz-QNet is capable of modeling far larger gene sets, as n genes require $n$ qubits for encoding, provided robust error-mitigation techniques are applied.}

\section{Simulations}
\label{results:disscuss}

Our work is integrated with Qiskit, an open-source quantum computing library that simulates a noisy quantum circuit using the Aer Simulator backend with default parameters.

\textcolor{black}{\subsection{Dataset Processing and Parameter Initialization}}

\textcolor{black}{To construct a Gene Regulatory Network (GRN) focused on entorhinal cortex pathology in Alzheimer’s disease (AD), we applied a three-stage filtering strategy that balances biological relevance, data robustness, and therapeutic potential to select a limited number of genes for this study amongst $100+$ gene candidates identified by the GWAS. First, we ensured EC-specific dysregulation by selecting genes that were both detectable and variable in single-nucleus RNA-seq data ($GSE138852$) from the entorhinal cortex, an early site of Alzheimer’s pathology; genes with stable expression and meaningful variability across nuclei are prioritized to strengthen model reliability. Next, we ensure pathway coverage by selecting genes that represent key processes implicated in Alzheimer’s disease, including amyloid processing ($APP$, $PLD3$)~\cite{selko2016, zhang2011}, lipid metabolism and insulin signaling ($SREBF2$, $AKT3$)~\cite{huang2012, barb2018}, and synaptic structure and function ($GAS7$, $FGF14$)~\cite{hsu2017,gotoh2013}, thereby capturing the interplay among multiple pathogenic mechanisms. Finally, we emphasize therapeutic actionability by prioritizing genes such as $EGR1$ and $YY1$~\cite{wein2017,nowak2006} that are known or emerging targets for Alzheimer’s interventions, enhancing the translational relevance of the predicted interactions. This multi-filter approach ensures that the resulting network reflects biologically significant, data-supported, and clinically actionable relationships. The limited number of genes is also chosen in accordance with quantum simulation constraints (hardware, errors, and scalability), and makes a strong proof-of-concept model.}\\
While gene expression is by definition a continuous process, we used binarization at the $0$ Pearson residual threshold specifically as a mean to augment both biological interpretability and computational tractability. The method accounts for technical noise characteristic of single-nucleus RNA sequencing, namely, dropout events, where low-expressed genes may not be detected due to finite $mRNA$ capture. With a zero threshold, we separate biologically significant expression (residual $> 0$) from possible technical arte-facts, adhering to best practices for sparse $scRNA-seq$ data processing as outlined in recent research.\\
Biologically, numerous Gene Regulatory Networks (GRNs) function through specialised activation mechanisms—gene silencing or transcription factor binding—where binary states of expression (expressed or not expressed) more accurately reflect switch-like behaviour. Given that our quantum regression model is formulated to model these discrete regulation events, binary input is both biologically appropriate and methodologically consistent with our objectives~\cite{grn2008}.\\
Furthermore, representing continuous values of gene expression in quantum circuits would require many more qubits to reach the same level of numerical accuracy, which is typically beyond what is possible with today's Noisy Intermediate-Scale Quantum (NISQ) technology. Consequently, binarization enables the implementation of quantum-efficient circuits with preservation of the underlying regulation rules required for simulations of interacting genes.\\
\textcolor{black}{The dataset GSE138852~\cite{jovic2022} comprises single-nucleus RNA sequencing ($snRNA-seq$) data from entorhinal cortex (EC) tissues of twelve individuals, including both control and Alzheimer’s disease (AD) brains. From this dataset, a total of $13,214$ high-quality nuclei are identified, all derived from AD-positive patients, and an initial subset of $1,041$ nuclei expressing at least one target gene was used for downstream Gene Regulatory Network (GRN) modeling. \\
All raw UMI counts were processed in RStudio using the Seurat and SCTransform packages. First, the raw expression matrix has been normalized using the Pearson residual method implemented in SCTransform, which corrects for sequencing depth and gene-specific technical biases while stabilizing variance across nuclei. The normalized residuals were then log-transformed and binarized at a threshold of zero: for each gene, residuals greater than zero are assigned a value of ‘$1$’ (expressed) and residuals equal to or below zero were assigned ‘$0$’ (non-expressed). \\
Binarization at the $0$ Pearson residual threshold is implemented for three primary reasons: (1) Biological appropriateness: gene regulatory events often exhibit switch-like behavior (activation/silencing) well-captured by binary states~\cite{grn2008}. Given that our quantum regression model is formulated to model these discrete regulation events, binary input is both biologically suitable and methodologically consistent with our objectives; (2) Technical necessity: it distinguishes true biological expression (residual $> 0$) from dropout artifacts in sparse $snRNA-seq$ data~\cite{jovic2022}; (3) Quantum tractability: continuous encoding would require prohibitive qubit resources on NISQ hardware. This approach aligns with the best recent practices for GRN inference from single-cell data, where discrete activation states preserve core regulatory logic.\\
This binarization step separates true biological signal from technical dropouts and produces a binary activation state for each gene in each nucleus. To ensure that the final input captured meaningful biological information, only nuclei with detectable expression of at least one of the eight selected target genes ($APP$, $PLD3$, $SREBF2$, $AKT3$, $EGR1$, $YY1$, $FGF14$, $GAS7$) are retained, yielding a final filtered dataset of $1,041$ nuclei. The resulting binarized expression matrix $X_b$ has dimensions $8 \times m$, where each row corresponds to a target gene and each column to a filtered nucleus. For quantum circuit input, each nucleus is encoded as an 8-bit binary string representing its gene activation state. The observed probability distribution $p_{obs}$ is calculated as the frequency of each unique binary label among all $m$ cells. Labels with zero occurrence are assigned zero probability, and the distribution is renormalized to sum to one, ensuring that only informative states contributed to GRN training. This approach mitigates sparsity and dropout artifacts common in single-cell data by focusing on nuclei that provide actionable regulatory information. Finally, the variational quantum circuit is initialized by setting the non-diagonal elements of the controlled-rotation gates ($C_{R_Y}$) to zero, while the diagonal rotation angles for each gene-specific qubit were initialized using $\theta = 2 \arcsin(\sqrt{a_k})$, where $a_k$ represents the activation ratio of the $k^{th}$ gene. This initialization guarantees that the probability of measuring an active (‘$1$’) state for each qubit aligns with the empirical activation ratio after the encoding layer $L_e$~\cite{roman2022}, ensuring that the model’s quantum state reflects the biological signal in the input data.}

\subsection{Simulation Results}

\begin{figure}[thbp]
	\centering
  \subcaptionbox{}{\includegraphics[scale=0.35]{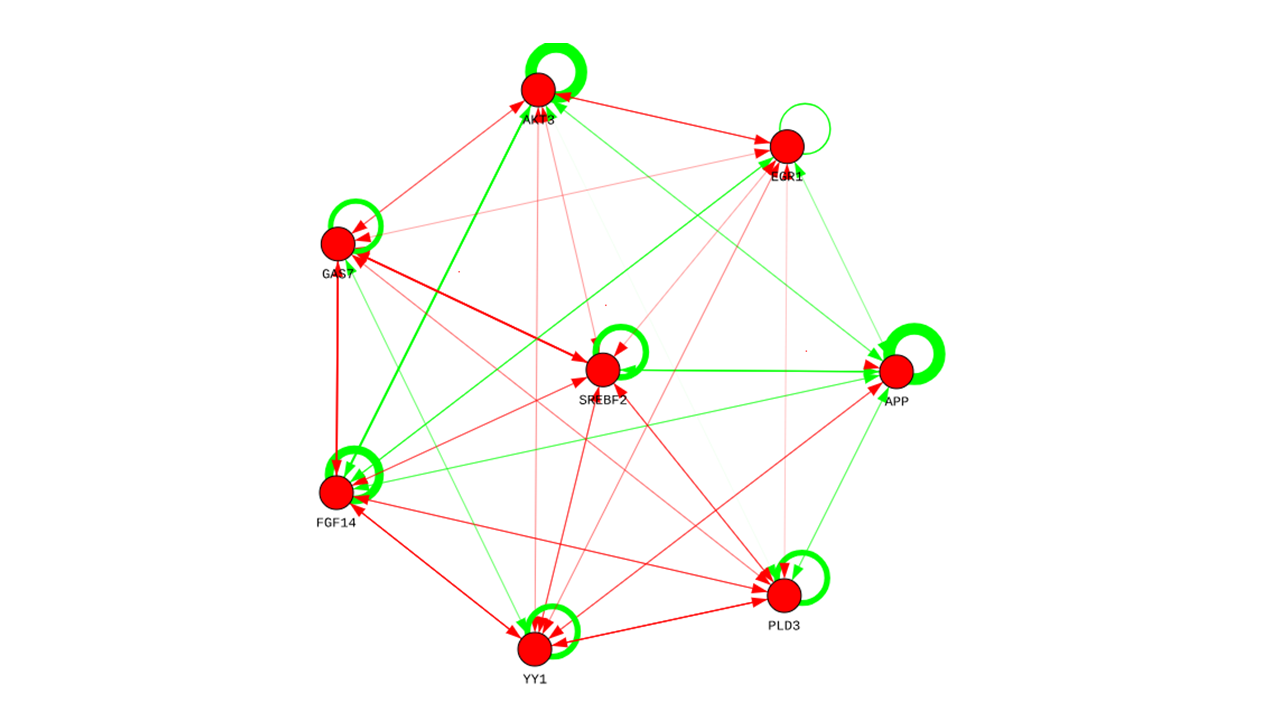}}
	\subcaptionbox{}{\includegraphics[scale=0.15]{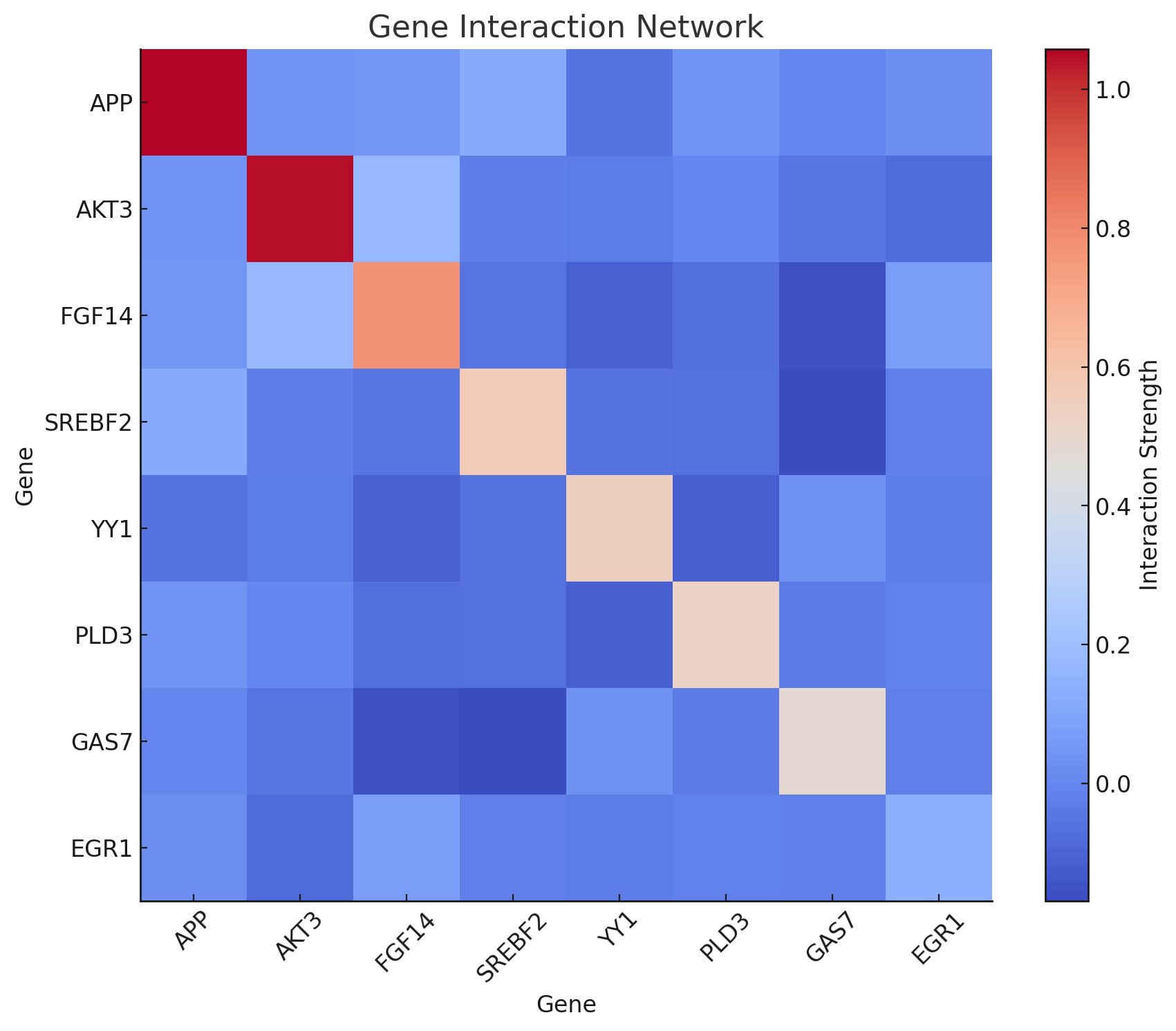}} 
    \caption{(a) The nodal graph, where each node corresponds to the specific gene and the edges determine the gene regulatory interactions. Green edges represent upregulation, and red edges represent downregulation. The weight of the edges determines the strength of the interaction between genes. (b) Heat map generated to study Alzheimer's gene interactions for our model and to quantify graph results.}
  \label{fig:QGRN_result}
\end{figure}

\begin{figure*}[htbp]
\vspace{-0.35in}
	\centering
	\subcaptionbox{}{\includegraphics[scale=0.22]{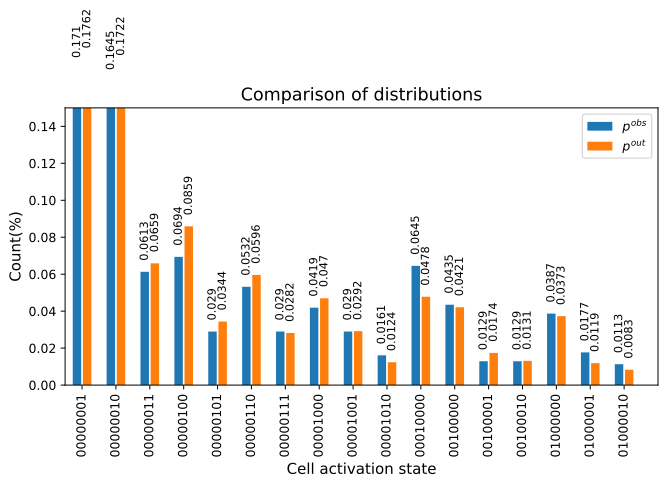}}
         \subcaptionbox{}{\includegraphics[scale=0.22]{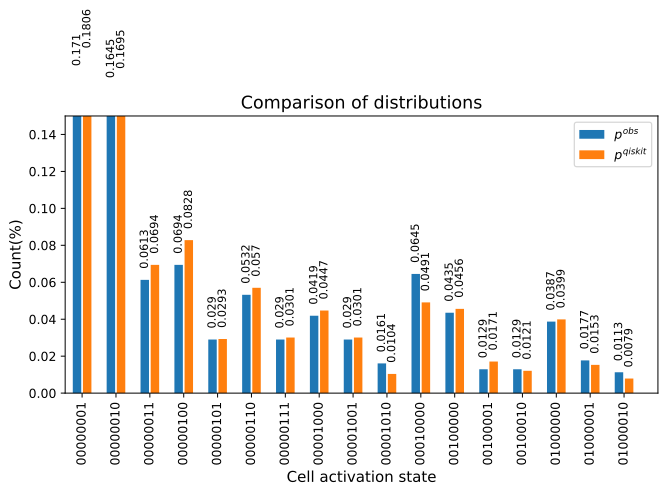}}
          \subcaptionbox{}{\includegraphics[scale=0.18]{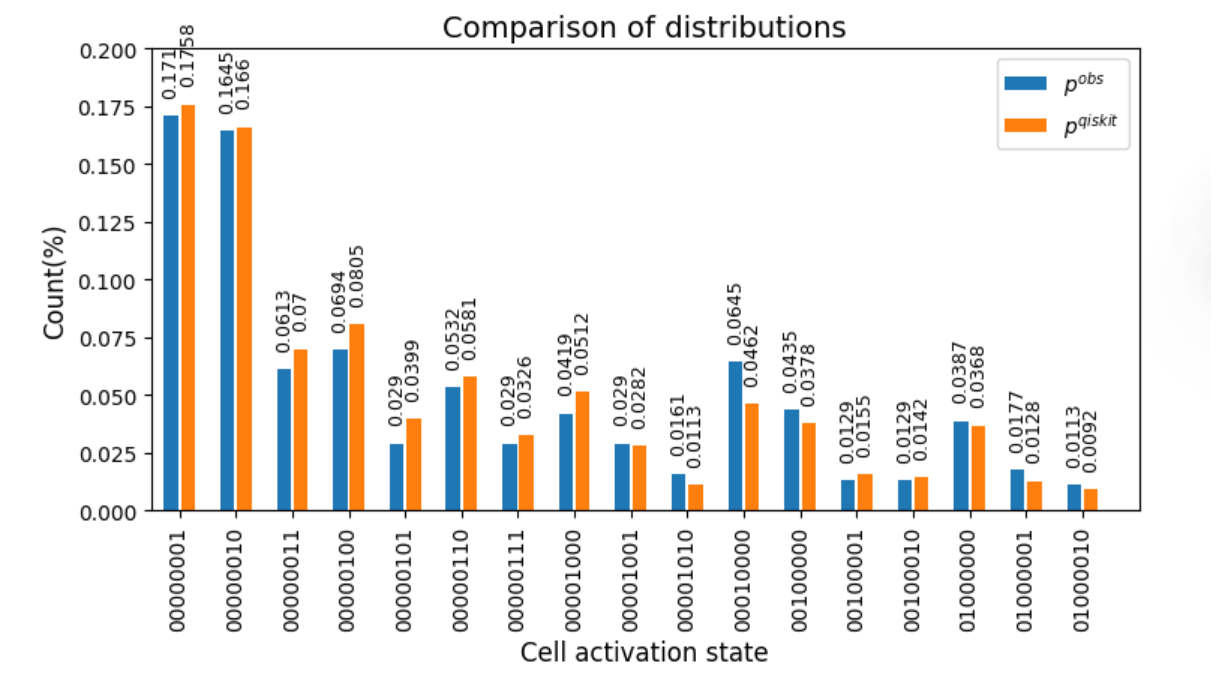}}
         %\subcaptionbox{}{\includegraphics[scale=0.1]{Heatmap (1).jpg}}
	\caption{(a) The observed vs output frequency distributions graph for Alz-QNet, (b) The observed vs simulated frequency distributions graph for Alz-QNet, and (c) {The observed vs simulated frequency distributions graph for QGRN}.} %(d) A quantum regression network to study Alzheimer's gene interactions.}
\label{fig:graph1}
\vspace{-0.1in}
\end{figure*}

In our investigation, we used $snRNA-seq$ EC data from AD patients to explore the genetic and molecular mechanisms underlying this neurodegenerative disorder ($GSE138852$). The EC, paramount for memory and navigation, is among the primary regions impacted by AD, making it an ideal focus for studying early pathological changes. From this dataset, we selected $8$ specific genes: $APP$, $SREBF2$, $GAS7$, $PLD3$, $YY1$, $FGF14$, $AKT3$, and $EGR1$. %and mapped according to Table~\ref{tab:gene_mapping}. 
In quantum simulations, a nodal graph is generated using Qiskit along with the observed vs. output frequency distribution graph and the observed vs. simulated frequency distribution graph, as shown in Fig.~\ref{fig:QGRN_result}, which resembles the classical GRN~\cite{david2005}. In addition, we also simulated the QGRN as a baseline model and found that it produced a similar probability distribution to Alz-QNet, as illustrated in Fig.~\ref{fig:graph1}, while saving computations.
%\vspace{-0.35in}
%\begin{equation}]
% \scalebox{0.55}{
% \begin{bmatrix}
%\scalebox{0.55}{
 $$\tiny\left[{{\begin{array}{cccccccc}
\textcolor{black}{1.058} & \textcolor{black}{0.044} & \textcolor{black}{0.050} & \textcolor{black}{0.123} & \textcolor{black}{-0.054} & \textcolor{black}{0.044} & \textcolor{black}{-0.000} & \textcolor{black}{0.025} \\
\textcolor{gray}{0.044} & \textcolor{black}{1.044} & \textcolor{black}{0.174} & \textcolor{black}{-0.022} & \textcolor{black}{-0.025} & \textcolor{black}{0.002} & \textcolor{black}{-0.047} & \textcolor{black}{-0.078} \\
\textcolor{gray}{0.050} & \textcolor{gray}{0.174} & \textcolor{black}{0.773} & \textcolor{black}{-0.050} & \textcolor{black}{-0.105} & \textcolor{black}{-0.066} & \textcolor{black}{-0.153} & \textcolor{black}{0.074} \\
\textcolor{gray}{0.123} & \textcolor{gray}{-0.022} & \textcolor{gray}{-0.050} & \textcolor{black}{0.562} & \textcolor{black}{-0.058} & \textcolor{black}{-0.060} & \textcolor{black}{-0.168} & \textcolor{black}{-0.019} \\
\textcolor{gray}{-0.054} & \textcolor{gray}{-0.025} & \textcolor{gray}{-0.105} & \textcolor{gray}{-0.058} & \textcolor{black}{0.540} & \textcolor{black}{-0.112} & \textcolor{black}{0.034} & \textcolor{black}{-0.029} \\
\textcolor{gray}{0.044} & \textcolor{gray}{0.002} & \textcolor{gray}{-0.066} & \textcolor{gray}{-0.060} & \textcolor{gray}{-0.112} & \textcolor{black}{0.524} & \textcolor{black}{-0.033} & \textcolor{black}{-0.011} \\
\textcolor{gray}{-0.000} & \textcolor{gray}{-0.047} & \textcolor{gray}{-0.153} & \textcolor{gray}{-0.168} & \textcolor{gray}{0.034} & \textcolor{gray}{-0.033} & \textcolor{black}{0.485} & \textcolor{black}{-0.019} \\
\textcolor{gray}{0.025} & \textcolor{gray}{-0.078} & \textcolor{gray}{0.074} & \textcolor{gray}{-0.019} & \textcolor{gray}{-0.029} & \textcolor{gray}{-0.011} & \textcolor{gray}{-0.019} & \textcolor{black}{0.138} \\
\end{array}}} \right]$$
\begin{figure*}
    \centering
    \includegraphics[width=\linewidth]{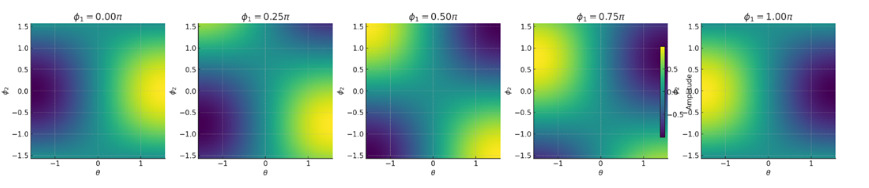}
    \caption{{Illustrates the impact of rotation angles $\theta$ and $\phi_1$ on the amplitude $|\mu|^2$ of the $|1\rangle$ state in a quantum circuit. Each heatmap corresponds to a different fixed value of $\phi_1$ ($0, 0.25\pi,0.5\pi,0.75\pi$), while $\theta$ (X-axis) and $\phi_1$ (Y-axis) vary continuously. The colour intensity represents the amplitude magnitude of the target qubit's $|1\rangle$ state, with brighter regions indicating higher amplitudes.}}
    \label{fig:heatmap}
\end{figure*}
% \end{bmatrix}}
% /]
%\end{equation} 
% \begin{table}[h!]
% \centering
% \begin{tabular}{|c|c|}
% \hline
% \textbf{Column/Row number} & \textbf{Gene mapped} \\ \hline
% 1 & $APP$ \\ \hline
% 2 & $AKT3$ \\ \hline
% 3 & $FGF14$ \\ \hline
% 4 & $SREBF2$ \\ \hline
% 5 & $YY1$ \\ \hline
% 6 & $PLD3$ \\ \hline
% 7 & $GAS7$  \\ \hline
% 8 & $EGR1$ \\ \hline
% \end{tabular}
% \caption{Gene Mapping}
% \label{tab:gene_mapping}
% \end{table}
The above matrix shows the $\theta_{x,y}$ values, where each row and column corresponds to a different gene. The black upper triangular matrix is the optimized value of theta obtained, and the lower triangular matrix was constructed from the equal values of $\theta_{x,y} = \theta_{y,x}$. \\
The red lines show the down-regulated networks, and the green lines show the up-regulated networks. Our graph model delves deeper into the genetic interactions of an AD patient's brain taken from the EC.
In addition to the gene interactions that affect prominent genes like $APP$ and $AKT3$, which have direct correlations, our findings extend and support the detailed epigenetic and transcriptomic insights~\cite{grubman2019}, particularly through the interactions involving $YY1$ and $PLD3$, as shown in Fig.~\ref{fig:QGRN_result}.\\
Fig.~\ref{fig:heatmap} illustrates the impact of rotation angles $\theta$ and $\phi_1$ on the amplitude $|\mu|^2$ of the $|1\rangle$ state in a quantum circuit. Each heatmap corresponds to a different fixed value of $\phi_1$ ($0, 0.25\pi,0.5\pi,0.75\pi$), while $\theta$ (X-axis) and $\phi_1$ (Y-axis) vary continuously.The heatmaps reveal symmetric patterns across the parameter space. This symmetry, stemming from the property $\theta_{x,y} = \theta_{y,x}$, allows us to encode the same amplitude information with half the number of gates. For example, the heatmap values for ($\phi_1, \theta$) and ($\phi_2, \theta$) by fixing either and varying the other, and observing the symmetry in the heatmap to conclude that they overlap significantly, demonstrating that redundant gates contribute no additional unique information. The regions of high and low amplitude values are nearly invariant to the interchange of $\phi_1$ and $\phi_2$, further validating the symmetric property of the $C_{R_Y}$ gates. Certain boundary conditions (e.g., $\phi_1=0$ or $\phi_1=\pi$) exhibit near-linear dependence of amplitudes, simplifying the overall amplitude structure. Leveraging these symmetric properties, we reduce the number of $C_{R_Y}$ gates by $50\%$ without altering the final quantum state representation. This reduction directly translates to lower computational cost and complexity, particularly for large QGRN networks with multiple gene-gene interactions.

\section{Discussions}
\label{sec:discus}

\textcolor{black}{In this article, integrating transcriptomic data, our Alz-QNet offers insights into how regulatory mechanisms contribute to AD}. This supports the study conducted by Grubman~\emph{et al.}~\cite{grubman2019} with an emphasis on considering transcriptional and epigenetic factors to understand brain function and disease, highlighting the importance of integrating multiple layers of biological data to understand the molecular foundations of diseases like Alzheimer's. This interdisciplinary methodology offers valuable insight into the molecular foundations of AD and underscores the promising role of QML research in elucidating complex biological phenomena. 

\subsection{$YY1$ and $PLD3$ Interaction}
$YY1$ is a transcription factor with dual gene activation and repression roles. It influences gene expression epigenetically by recruiting proteins that modify chromatin structure. Our results show a negative interaction between $YY1$ and $PLD3$ ($-0.112907$), suggesting $YY1$'s repressive role on $PLD3$. This aligns with the known function of $YY1$ in gene repression and epigenetic regulation. Our model forecasts $YY1$ to repress $PLD3$, in line with established epigenetic mechanisms. $YY1$ recruits HDACs and DNA methyltransferases to condense chromatin at $PLD3$, repressing its expression. Research documents YY1-dependent hypermethylation of the $PLD3$ promoter in neurodegenerative settings. The predicted YY1-$PLD3$ interaction intensity ($\theta = -0.112$) infers a repressive relationship, and therefore, $PLD3$'s lysosomal activity in AD could be under epigenetic control. Our quantum model highlights transcriptional networks and indicates the direction of integrated epigenetic-transcriptional therapies, e.g., HDAC inhibitors, derepressing $PLD3$ without interfering with $YY1$’s regulation of targets such as $SREBF2$.

\subsection{Role of $PLD3$ in AD}
$PLD3$ is involved in processing $APP$ and regulating amyloid-beta levels, which are critical in AD pathology. Our study reveals complex interactions between $PLD3$ and other genes, such as $APP$ and $SREBF2$. These interactions suggest that the regulatory network of $PLD3$ is influenced by both transcriptional and epigenetic mechanisms, supporting the emphasis of the neurobiology literature on the importance of epigenetic modifications in brain development and disease.

\textcolor{black}{\subsection{Alz-QNet Model Predictions}}

\textcolor{black}{Our quantum model, Alz-QNet, predicts a robust network of six core gene interactions driving Alzheimer’s disease pathogenesis, many of which are experimentally supported by independent studies (Supplementary Table S1). The strong alignment between our predicted $\theta$-values and established mechanisms, such as $APP$ activating $SREBF2$~\cite{barbero2013}, $FGF14$ repressing $SREBF2$~\cite{hsu2017}, and $PLD3$ repressing $APP$~\cite{bottero2021} demonstrates that Alz-QNet can recover biologically meaningful regulatory circuits. Notably, our nodal graph analysis reveals $PLD3$ as a key repressor influencing not only $APP$ but also $GAS7$ and $SREBF2$, linking amyloid processing with lipid metabolism and synaptic stability. The model further suggests that $YY1$ inhibits $PLD3$, creating an indirect repression loop over $SREBF2$, while $EGR1$ emerges as a major transcriptional activator for $APP$, $SREBF2$, and $GAS7$. These regulatory patterns align with experimental findings showing, for example, that $EGR1$ knockdown can reduce tau phosphorylation and improve cognition in AD models, and that $YY1$ recruits epigenetic repressors like PRC2 and DNMTs to silence targets such as $PLD3$~\cite{wein2017,grubman2019}. Additional pathways predicted by Alz-QNet, including the $AKT3$–$FGF14$–$GAS7$ axis, highlight the dual role of $AKT3$ in promoting neuronal survival but also contributing to insulin resistance, a known AD risk factor. This interconnected network suggests feedback loops coordinating amyloid metabolism, lipid dysregulation, insulin signaling, and synaptic maintenance. \\
Nodal analysis reveals the $APP$–$PLD3$ axis as a central regulatory hub linking amyloid processing and lipid metabolism, with $PLD3$ repression of $APP$ and $SREBF2$ indicating dual functional roles. Alz-QNet’s predictions prioritize therapeutically actionable nodes: (i) inhibiting $YY1$ (e.g., with resveratrol analogs) to derepress $PLD3$ and enhance lysosomal clearance; (ii) suppressing $EGR1$ (e.g., via $CRISPRi$) to reduce $APP$ and $SREBF2$ expression, lowering amyloid burden and lipid dysregulation; and (iii) activating $PPAR\gamma$ (e.g., with pioglitazone) to modulate $FGF14$, particularly benefiting insulin-resistant AD subgroups. By mapping these feedback loops and gene-gene interactions, Alz-QNet moves beyond amyloid-centric approaches, offering testable, multi-target strategies that align with known pathways and expand therapeutic avenues for AD, which could be useful in biomarker discovery and gene therapy-based therapeutics.}\\
\textcolor{black}{We propose three scalability strategies: (1) Subnetwork partitioning to focus on correlated gene clusters; (2) Tensor-ring compression to reduce parameter space; and (3) Classical preprocessing for dimensionality reduction. Hybrid quantum-classical workflows will enable efficient scaling while mitigating noise. With further evolution and improvements in the field of quantum noise correction and by using approximatory techniques, large gene dataset studies using $sc/snRNA$ could be computationally feasible in the near future. However, issues regarding quantum noise will still come into play, and using the above proposed techniques of error mitigation could be the solution to have improvements in the accuracy of large data processing with improved accuracy in quantum hardware in the near future.\\
While our findings were obtained using noise-free simulation (Qiskit Aer), current quantum processors face limitations including small qubit counts (typically $20-50$ usable qubits), limited coherence times ($50-200$ µs), and two-qubit gate error rates of $1-2\%$. Our symmetry-based gate reduction, halving the number of controlled-rotation $(C_{R_Y})$ gates, directly mitigates these challenges by reducing circuit depth and two-qubit operations. For near-term applicability, Alz-QNet is compatible with error-mitigation techniques including zero-noise extrapolation, probabilistic error cancellation, tensor-ring approximations, and measurement error correction. Each of these error mitigation methods can significantly enhance output fidelity on quantum hardware constructed by IBM Quantum, IonQ, or Rigetti by $2\times$ to $5\times$. Hardware-aware compilation (e.g., qubit mapping optimizations) and hybrid quantum-classical protocols (where quantum circuits learn GRN parameters while classical resources handle preprocessing and noise correction) provide a clear path toward experimental demonstration. In our framework, each gene is mapped to a single qubit, meaning that an $n$-gene network requires $n$ qubits. While full hardware validation remains future work, our gate optimization enables modelling of $50$-gene networks on current $50$-qubit devices. As quantum processors scale beyond $1,000$ qubits, Alz-QNet will support larger gene sets with robust error mitigation}.\\

\section{Conclusion}
\label{sec:concl}

The current study proposes a novel approach that combines the principles of QML and GRN analysis to investigate the complex gene-gene interactions involved in AD pathology within the EC. Our proposed Alz-QNet model has the potential for significantly enhanced computational power, enabling faster and more precise modeling of intricate gene interactions. This advancement can lead to a deeper understanding of biological processes and can offer groundbreaking insights into gene regulation and expression. Leveraging the precision of quantum computing could revolutionize personalized medicine by tailoring treatments to individual genetic profiles. Furthermore, quantum algorithms have the capability to optimize biological processes and provide efficient solutions to complex biological challenges, making them a potent tool for simulating and understanding intricate systems. Notably, we have reduced the computational cost by nearly half compared to traditional QGRN, providing a progressive edge in addressing the computational expense issue.\\
However, the current state of quantum computing is still in its nascent stages, with limited availability and challenges such as errors and decoherence. Modeling the GRN at the quantum level is highly complex and resource-intensive, necessitating specialized hardware, software, and expertise, which can be costly. Scalability poses a significant hurdle, as existing quantum computers struggle to handle the vast datasets typical of gene regulatory networks. Given the constraints of hardware qubits and noise, approximation techniques such as the utilization of tensor rings may offer a viable approach in the future for scaling up data input while ensuring accurate results.

\section{Data and Code availability}
\label{data}
 The Alzheimer's dataset used for our model can be found at: \url{https://www.ncbi.nlm.nih.gov/geo/query/acc.cgi?acc=GSE138852}. The Pytorch code for our Alz-QNet implementation is available in Qiskit simulations at \url{https://github.com/NeeravSreekumar/AD_Quantum}.

\bibliographystyle{elsarticle-harv} 
\bibliography{refs}
%\bibliography{Deb}
%\pagebreak
\onecolumn
\section*{Supplementary Information}

\begin{table}[h]
\centering
\caption{\textcolor{black}{Supplementary Table S1: Literature support for all predicted interactions}}
\label{tab:s1}
\begin{tabular}{|c|c|p{2cm}|p{10cm}|}
\hline
\textbf{Interaction} & \textbf{Direction} & \textbf{Validated?} & \textbf{Supporting Evidence} \\ \hline
APP $\rightarrow$ SREBF2 & Activation & Yes & APP/PS1 mice overexpressing SREBF2 show accelerated A$\beta$/tau pathology (Barbero-Camps et al., 2013) \\ \hline
FGF14 $\rightarrow$ SREBF2 & Repression & Yes & PPAR$\gamma$ agonists (activating FGF14) reduce SREBF2 activity (Hsu et al., 2017) \\ \hline
YY1 $\rightarrow$ SREBF2 & Repression & Yes & YY1 haploinsufficiency upregulates SREBF2 in neurons (Pan et al., 2021) \\ \hline
EGR1 $\rightarrow$ APP & Activation & Yes & EGR1 binds APP promoter and increases amyloid production (Qin et al., 2016) \\ \hline
PLD3 $\rightarrow$ APP & Repression & Yes & PLD3 knockout increases APP processing and A$\beta$ plaques (Bottero et al., 2021) \\ \hline
YY1 $\rightarrow$ EGR1 & Repression & Yes, but dependent & YY1 loss upregulates EGR1 in progenitors but not mature neurons (Periera et al., 2025) \\ \hline
\end{tabular}
\end{table}
\end{document}